\documentclass[journal=janac3,manuscript=article, layout=twocolumn]{achemso}

\usepackage[version=3]{mhchem}

\usepackage{graphicx}
\usepackage{verbatim}
\usepackage{textcomp,bbm}
\usepackage[breaklinks=true,colorlinks=true,linkcolor=black,urlcolor=black,citecolor=black]{hyperref}
\usepackage{lineno}

\usepackage{amsmath}
\usepackage{graphicx}
\usepackage[skip=6pt, indent=0pt]{parskip}
\usepackage{caption}
\usepackage{subcaption}
\usepackage{multirow}
\usepackage{hyperref}
\usepackage{bibentry}
\usepackage{hyphenat}
\usepackage{float}
\usepackage{color}
\usepackage{soul}

\title{Accurate single-nanoparticle sizing down to 3~nm with an optofluidic microcavity}

\author{Shalom Palkhivala}
\affiliation{Karlsruher Institut f\"ur Technologie, Physikalisches Institut, Wolfgang-Gaede-Str. 1, 76131 Karlsruhe, Germany}

\author{Larissa Kohler}
\affiliation{Karlsruher Institut f\"ur Technologie, Physikalisches Institut, Wolfgang-Gaede-Str. 1, 76131 Karlsruhe, Germany}

\author{Christian Ritschel}
\affiliation{Karlsruher Institut f\"ur Technologie, Institut f\"ur Anorganische Chemie, Engesserstr. 15, 76131 Karlsruhe, Germany}

\author{Claus Feldmann}
\affiliation{Karlsruher Institut f\"ur Technologie, Institut f\"ur Anorganische Chemie, Engesserstr. 15, 76131 Karlsruhe, Germany}

\author{David Hunger}
\affiliation{Karlsruher Institut f\"ur Technologie, Physikalisches Institut, Wolfgang-Gaede-Str. 1, 76131 Karlsruhe, Germany}
\alsoaffiliation{Karlsruher Institut f\"ur Technologie, Institut f{\"u}r QuantenMaterialien und Technologien , Hermann-von-Helmholtz-Platz 1, 76344 Eggenstein-Leopoldshafen, Germany}
\email{david.hunger@kit.edu}

\begin{document}

\begin{abstract}
Nanoparticles are ubiquitous, and methods that reveal insights into single-particle properties are highly desired to enable their advanced characterization. Techniques that achieve label-free single-nanoparticle detection often lack bandwidth or do not provide quantitative information. Here, we present a cavity-based dispersive sensing method that achieves a high bandwidth to capture all relevant timescales of translational diffusion, and a sensitivity to detect and size single particles with diameters down to 3~nm. We develop an analytical model describing the autocorrelation function for particle diffusion in a standing-wave sensing geometry and propose a method to address the challenges posed by the transient nature of single-particle signals.
With this, we achieve quantitative particle sizing with high precision and accuracy, and provide an important tool to analyze single-particle diffusion. 

\textbf{Keywords:} single-particle detection, Brownian motion, autocorrelation, sizing, optical microcavity, gold nanoparticles.
\end{abstract}

\section{Introduction}
Nanomaterials are often heterogeneous, and methods that give direct access to single particle properties can open up the view into the details of the distribution, reveal features that are otherwise washed out in ensembles, and enable the studies of dynamical behaviour. Therefore, a range of label-free single-particle sensing techniques has been developed. For example, interferometric scattering microscopy (iSCAT) \cite{Taylor2019,Ginsberg2025} has enabled the detection of single nanoparticles and biomolecules \cite{Piliarik2014} with sizes below 10~kDa using machine learning algorithms \cite{dahmardeh}, and the tracking \cite{kazaian} or sizing \cite{kashnakova2022} of single nanoparticles down to 10~nm in diameter.
 
Another approach is the use of nanoplasmonic hot spots, e.g. at the tip of gold nanorods \cite{Baaske2020,Asgari2023}, or whispering gallery mode (WGM) resonators with high quality factors such as microspheres \cite{Vollmer2012,Foreman2015}, which enable real-time detection of single particles as well as quantitative particle sizing \cite{zhu2010}. Very high sensitivity is achievable by combining WGM cavities with nanoplasmonic hot spots \cite{Baaske2014}.
However, most of these sensors are based on the interaction of the analyte with the near-field and thus the sensor surface, which may constrain and modify diffusion. This can preclude a precise quantification of, for example, the hydrodynamic radius. Open access microcavities avoid this limitation \cite{Trichet2016,Smith2022,Kohler2021, Malmir2022}, but the larger mode volume compared to nanoplasmonic sensors has so far restricted detection and sizing to particles $>~100$~nm. Only recently, open cavities could achieve sensitivities for single molecules down to $\sim 1$~kDa using a highly non-linear photothermal regime \cite{needham}. Although this remarkable sensitivity has allowed for single protein detection, extracting information relating to the size or shape with this approach has remained limited to the statistics obtained from many detection events. In addition, the influence of the photothermal response that convolves with the signal makes quantitative and unbiased evaluation challenging.

Here, we operate a fiber Fabry-Perot microcavity locked on the slope of a resonance in the linear dispersive regime. Using a high finesse of around 55,000 and a high stability with cavity length fluctuations of about 300~fm, we achieve a sensitivity capable of detecting individual 3~nm gold nanoparticles that diffuse through the cavity mode. With a detection bandwidth that is in principle limited only by the decay rate of the cavity ($\sim$~500~MHz), we can obtain sufficient statistics from single-particle transits through the cavity mode to calculate the autocorrelation function (ACF) for further analysis.

While such an analysis is well established for ensemble techniques such as fluorescence correlation spectroscopy (FCS) and dynamic light scattering (DLS) \cite{wohland}, the scenario studied here raises two central aspects that require careful attention: First, the probing volume in FCS is well approximated by a 3D Gaussian that can be treated with an analytical model for the ACF, while a cavity standing wave has a more complex structure that requires a reformulation of this model. Second, the extraction of diffusion time constants and hydrodynamic radii from the ACF relies on the assumption of a stationary system, which can be satisfied by an ensemble with a constant average concentration. In contrast, single particle events are intrinsically non-stationary. We address both aspects and propose an analytical model for the ACF of particle diffusion through a standing wave probe volume, and introduce an analysis that minimizes undesired influence of the transient single-particle behavior. With this methodology, we perform quantitative particle sizing for a range of nominal particle diameters and obtain size distributions that are in good agreement with reference measurements performed with DLS and transmission electron microscopy (TEM). This opens the way to quantitatively assess nanomaterials such as unlabeled biomolecules in their native environment, and could allow further dynamic processes such as rotation and conformational changes to be investigated in the future.

\begin{figure}
  %  \centering
    \includegraphics[width=0.45\textwidth]{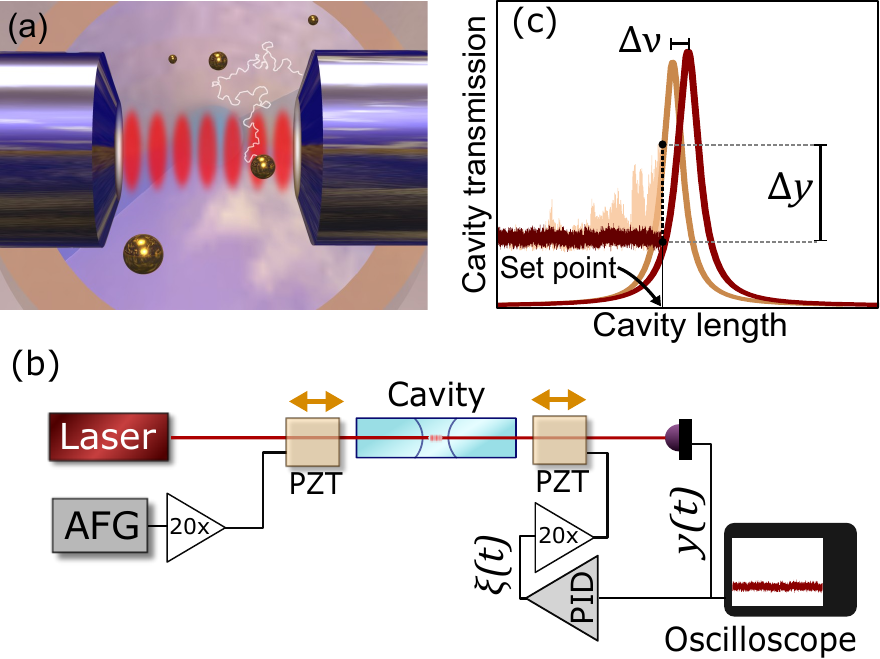}
    \caption{(a) Illustration of gold nanospheres diffusing through an optofluidic fiber Fabry-Perot microcavity. (b) The optoelectronic setup for active cavity stabilization. The two cavity fibers are inserted into a glass ferrule with a lateral microfluidic channel. Piezoelectric transducers (PZT) driven by an arbitrary function generator (AFG) and proportional-integral-derivative (PID) feedback controller allow to tune the cavity mirror separation and to actively stabilize the cavity on the slope of a resonance respectively. For a detailed description, see Methods. (c) Detection of dispersive resonance shifts induced by nanoparticles via changes in transmission of a locked cavity.}
   \label{fig:setup}
\end{figure}

\section{Results}

\subsection{The locked optofluidic microcavity}
\label{sec:locked_cavity}

Fabry-Perot microcavities with high finesse strongly enhance the interaction between light and matter in the cavity mode volume, and therefore are powerful tools for the label-free detection of single nanoparticles in aqueous dispersion. Similarly to previous work \cite{Kohler2021}, we use a fiber-based optofluidic microcavity comprising two single-mode optical fibers inserted into a fiber-optic ferrule, with a microfluidic channel drilled into the ferrule perpendicular to the fibers, such that the cavity between the two fiber tips is immersed in water, see Fig.~\ref{fig:setup} (a), (b). The optical fibers are processed by laser machining to produce concave profiles with radii of curvature between 20~µm and 80~µm, and then coated with a highly reflective Bragg mirror having transmission $\mathcal{T} =$ 20~ppm in air \cite{hunger}.

We implement an actively stabilized cavity locked on the side of a resonance using a side-of-fringe-lock with an FPGA-based feedback circuit with a unity-gain bandwidth of 80~Hz, which is well below the frequency range covered by nanoparticle diffusion (see Supporting Information, Figs. S3 and S4). In comparison to previously reported length-modulated cavities \cite{Kohler2021,Trichet2016}, the locked cavity offers the possibility of measurements at much higher bandwidth, limited essentially only by the cavity linewidth. 

The cavity lengths used in our experiments were 3.7 -- 4.5~µm and the cavity finesse in water was ${4.5\times 10^4} - {5.8\times 10^4}$. Under active stabilization, the microcavity shows a root-mean-square optical length instability of $\sim$ 300 fm, which corresponds to about 5\% of the linewidth, over several hours. The intracavity power circulating in the locked cavity is in the order of 50 mW. A cavity-locked transmission trace is illustrated in Fig.~\ref{fig:setup}. We note that even the passively stable cavity shows high stability on short timescales (up to seconds), providing a low noise background for the measurements.

\subsection{Detection of single nanoparticles}
\label{sec:nanoparticle_detection}

A small particle with polarizability $\alpha$ diffusing though the optical field produces a relative resonance frequency shift of the cavity of ${\Delta\nu/\nu  = -\Re(\alpha)U^2(\mathbf{r_0})/2\varepsilon_\mathrm{m} V_{\mathrm{m}}}$, where $U(\mathbf{r_0})$ is the value of the normalized optical field distribution at the position $\mathbf{r_0}$ of the nanoparticle, $V_\mathrm{m}$ is the cavity mode volume and $\varepsilon_\mathrm{m}$ is the relative electric permittivity of the surrounding medium. For spherical particles in the Rayleigh regime with radius $R$ and relative permittivity  $\varepsilon_\mathrm{p}$, the polarizability is given by ${\alpha =4\pi\varepsilon_0 R^3(\varepsilon_\mathrm{p}-\varepsilon_\mathrm{m})/(\varepsilon_\mathrm{p}+2\varepsilon_\mathrm{m})}$. Small dispersive shifts $\Delta\nu$ produced by nanoparticles in the mode volume are translated into measurable changes $\Delta y$ in cavity transmission (Fig.~\ref{fig:setup}~(c)).

\begin{figure*}[ht!]
    \includegraphics[width=1\textwidth]{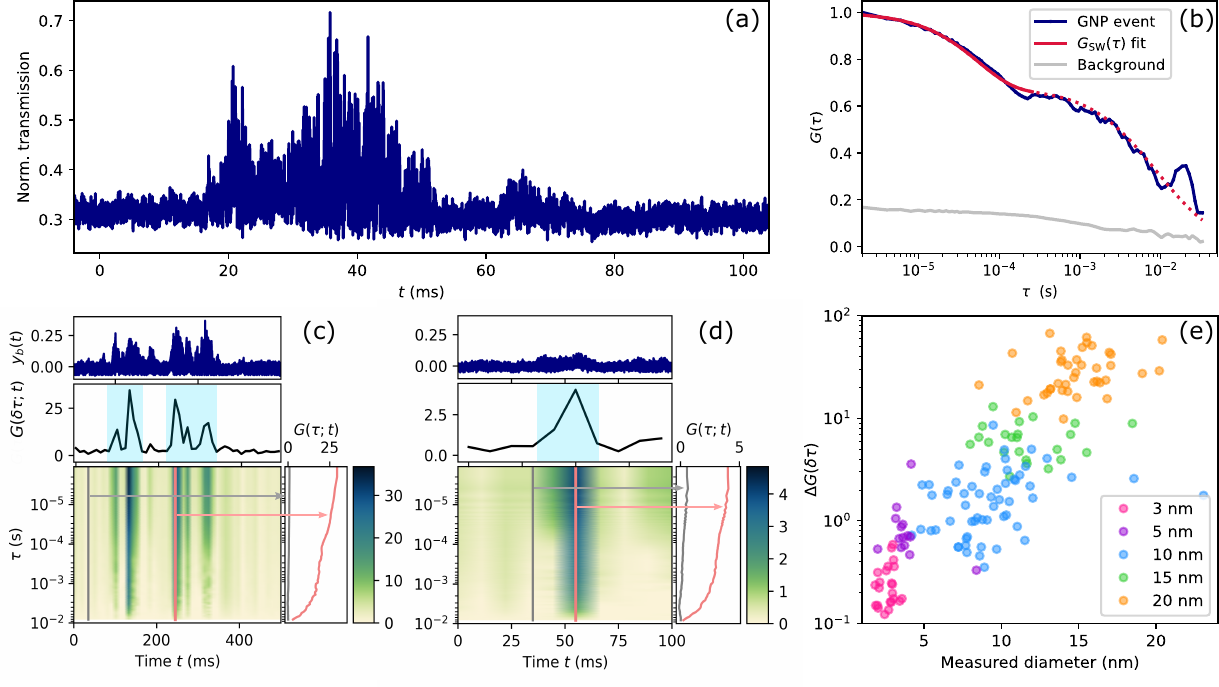}
    \caption{(a) Example of the transmission of a locked cavity disturbed by the diffusion of a 20~nm nanosphere through the optical mode. (b) The autocorrelation of the nanoparticle signal and its fit to the analytical ACF developed in this work, along with the relatively flat background autocorrelation. (c)--(d) Time traces (top panel), correlation amplitudes $G(\delta\tau)$ (middle panel), and autocorrelograms (bottom panel) of 20 nm and 10 nm nanosphere events respectively. (e) Each nanoparticle can be mapped onto the two-dimensional space of nanoparticle diameter and autocorrelation amplitude contrast $\Delta G(\delta\tau)$.}
    \label{fig:autocorrelation}
\end{figure*}

We use the microcavity to detect single citrate-stabilized spherical gold nanoparticles (GNP) in aqueous dispersion with specified diameters ranging from 3 nm to 20 nm. The stock sample was diluted to a concentration of 50 pM, so that there is, on average, less than one particle in the cavity mode at any time. The resulting suspension is injected into the microcavity, and the cavity transmission $y(t)$ is monitored with an avalanche photodiode on a 12-bit digital storage oscilloscope. %A ``nanoparticle event'' is said to occur when a nanoparticle transits through the optical field. 
Examples of single nanoparticle events are shown in Fig.~\ref{fig:autocorrelation}~(a) and the top panels of (c) and (d), as well as Fig. S5 in the Supporting Information.

We calculate the ACF $G(\tau)$ of the measured signal and observe characteristic decays of the correlation on two separated time scales, see Fig.~\ref{fig:autocorrelation}~(b). This differs from the autocorrelation typically observed by other methods when only translational diffusion is present \cite{Ries2012}, and we now analyze and model this behavior in detail to enable a fit of the data and extract physical quantities of interest, such as the hydrodynamic radius. As a first step, we use the autocorrelation to enable the detection of nanoparticles, even when the signal-to-noise ratio of the measured cavity transmission time trace does not allow a direct identification of nanoparticle events. For this purpose, we compute the autocorrelation of short segments of the data with duration $T_\mathrm{seg}$~=~10~ms, and evaluate the correlation amplitude $G(\delta\tau)$, where $\delta \tau$ is the shortest temporal delay in the autocorrelation, for each segment. The middle panels of Fig.~\ref{fig:autocorrelation}~(c) and (d) show examples of $G(\delta\tau)$ for 20~nm and 10~nm particles, respectively, while the lower panel shows the merging of ACFs of the segments into a time-resolved autocorrelogram. Whenever a particle is present, $G(\delta\tau)$ shows a value significantly larger than without a particle, since the empty-cavity background signal is largely uncorrelated. We thus use the autocorrelation amplitude contrast $\Delta G(\delta\tau)=G_\mathrm{NP}(\delta\tau)-G_\mathrm{BG}(\delta\tau)$ between the peak correlation amplitude $G_\mathrm{NP}(\delta\tau)$ of such a segment and the background correlation amplitude, $G_\mathrm{BG}(\delta\tau)$, as a measure of particle presence. The statistical distribution of $\Delta G(\delta\tau)$ for GNP events %(with peak correlation $G_\mathrm{NP}(\delta\tau)$) 
with different particle sizes is shown in Fig.~\ref{fig:autocorrelation}~(e), where a roughly exponential correlation of $\Delta G(\delta\tau)$ with particle size can be observed. This shows that $\Delta G(\delta\tau)$ alone already allows to differentiate particles of different size, but only qualitatively and with comparably large uncertainty. We note that the background value $G_\mathrm{BG}(\delta\tau)$ for times where no particle is present changes slightly between measurements due to changing noise levels (see Supporting Information, Fig. S5 (f)).  
However, the background remains a factor of $\sim 2$ smaller even for 3~nm GNPs. The use of the signal correlation contrast is thus a powerful way to detect very small particles whose signal would otherwise remain masked by noise. While it enables highly sensitive particle detection, provides qualitative size information, and does not require any modeling of the ACF, it does not provide a quantitative characterization of the particles. In addition, the magnitude of $\Delta G(\delta\tau)$ is dependent on parameters of our measurement system, such as cavity geometry, measurement bandwidth and the background, and hence does not allow the determination of nanoparticle properties without \textit{a priori} knowledge or calibration measurements. We therefore proceed to develop a proper model to enable accurate particle sizing.

\subsection{The autocorrelation function of diffusion in a standing wave field}
\label{sec:autocorrelation}
Autocorrelation-based analysis of diffusional motion is a widely used technique to determine the size of particles in suspensions by analyzing their diffusional motion, for instance in FCS \cite{Ries2012,elson,Dominguez2016}. For the accurate determination of time constants of diffusion, it is imperative to first determine the theoretical ACF for the optical mode geometry involved.

A typical FCS setup has a Gaussian focal spot which is approximated by the normalized intensity profile $W_\mathrm{FCS}(x,y,z)=\exp{(-2x^2/w_0^2)} \exp{(-2y^2/w_0^2)}\exp{(-2z^2/z_0^2)}$, where $2w_0$ is the focus diameter and $2z_0$ the depth of focus. The normalized ACF of particles with diffusivity $D_T$ is then
\begin{equation}\label{g_fcs}
    G_\textrm{FCS}(\tau)=\left(1+\frac{4D_T\tau}{w_0^2}\right)^{-1}\left(1+\frac{4D_T\tau}{z_0^2}\right)^{-1/2}.
\end{equation}

In FCS, the fluorescence $F(t)$ of particles diffusing through the focal spot of the excitation laser is measured, which is proportional to the time-dependent concentration $C(t)$ of fluorophores in the focal volume. Since the average concentration of particles is $\langle C(t)\rangle\gg1$, the fluctuation signal can be extracted from the measured fluorescence, $\delta F(t) = F(t) - \langle F(t)\rangle$ where $\langle F(t)\rangle$ is the steady-state average fluorescence signal. Fitting the autocorrelation of $\delta F(t)$ to $G_\textrm{FCS}(\tau)$ (Eqn. \ref{g_fcs}) allows the determination of the translational diffusion coefficient $D_T$. This directly yields information about the average hydrodynamic radius of the particles via the Stokes-Einstein relation $D_T=k_B T/(6\pi\eta R)$, where $\eta$ is the viscosity of the surrounding fluid and $T$ is its temperature.

In a Fabry-Perot cavity, the normalized intensity has a distribution given (up to the slowly-varying beam waist) by $W_\textrm{FP}(x,y,z)=\exp{(-2(x^2+y^2)/w_0^2)} \cos^2(kz)$, where $k=2\pi/\lambda$, and the autocorrelation integral has no analytical solution. Instead, a numerical ACF for a standing wave field can be calculated. Since an analytical function is essential for a fit-based analysis of the autocorrelation of particle diffusion, we propose an approximation of the analytical ACF by replacing the standing wave antinodes by a series of Gaussians separated by the standing wave period $\lambda/2$,
\begin{equation}\label{gauss_train}
    \cos^2(kz) \rightarrow  \sum_{m=-q}^{q}{e^{-s(kz-\frac{mk\lambda}{2})^2}}   \qquad q\rightarrow\infty.
\end{equation}

\begin{figure}[t!]
    \includegraphics[width=0.5\textwidth]{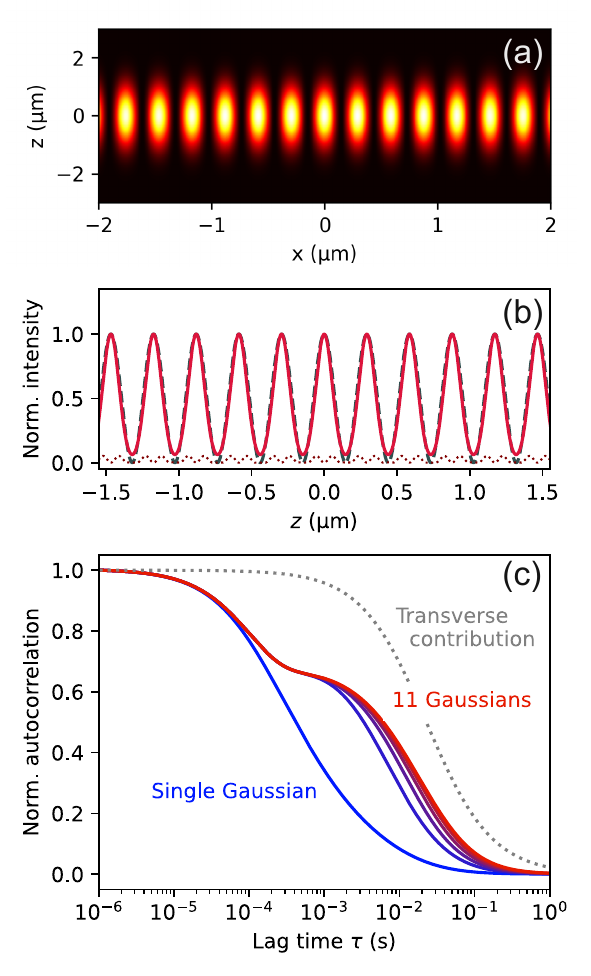}
    \caption{(a) Calculated normalized intensity of the cavity mode field. (b) The standing wave is modeled as a series of Gaussian functions with separation $\lambda/2$. The dashed line shows the deviation from the standing wave weighted with the normalized intensity. (c) The analytical ACF for diffusion through an optical field with varying numbers of Gaussians. The case of a single Gaussian corresponds to the ACF used in FCS.}
    \label{fig:gaussian_train}
\end{figure}

The scaling factor $s=2(\cos(1/e))^{-2}$ is introduced to match the $1/e^2$  levels of the two functions. Taking into account a finite number of antinodes, $q<\infty$, allows the ACF to be expressed as a closed-form analytical function 
\begin{equation}\label{g_sw}
    G_\textrm{SW}(\tau)=\frac{\sqrt{2k^2s}}{b_0\sqrt{\pi x}} e^{-\frac{a_0}{x}} \left(1+\sum_{m=1}^{2q}{b_m e^\frac{a_m}{x}}\right)
\end{equation}
where $x=1+2D_T k^2 s\tau$, and $a_m, b_m$ are analytically determined constants.  In Fig.~\ref{fig:gaussian_train}~(c), we compare the ACFs arising from different numbers of Gaussian antinodes used. We note that when $q=1$, the ACF corresponds to a single three-dimensional Gaussian focus and is identical to that used in FCS (Eqn.~\ref{g_fcs}). When  $q>1$, a second decay appears, which arises from diffusion through to multiple antinodes, i.e. due to the periodicity of the field seen by the nanoparticle. Additionally, the ACF converges as $q\gg1$, justifying the simplification we used to obtain Eqn.~\ref{g_sw}.
Furthermore, the transverse dimension of the cavity mode leads to a third decay time constant, shown as a grey dotted line in Fig.~\ref{fig:gaussian_train}. However, it overlaps with the multiple-antinode decay for our cavity geometry, and cannot be resolved. We verify that the approximations introduced do not cause systematic errors in subsequent ACF evaluation by Monte Carlo simulations (see Supporting Information, Fig. S1).

\begin{figure*}[ht!]
        \includegraphics[width=0.8\textwidth]{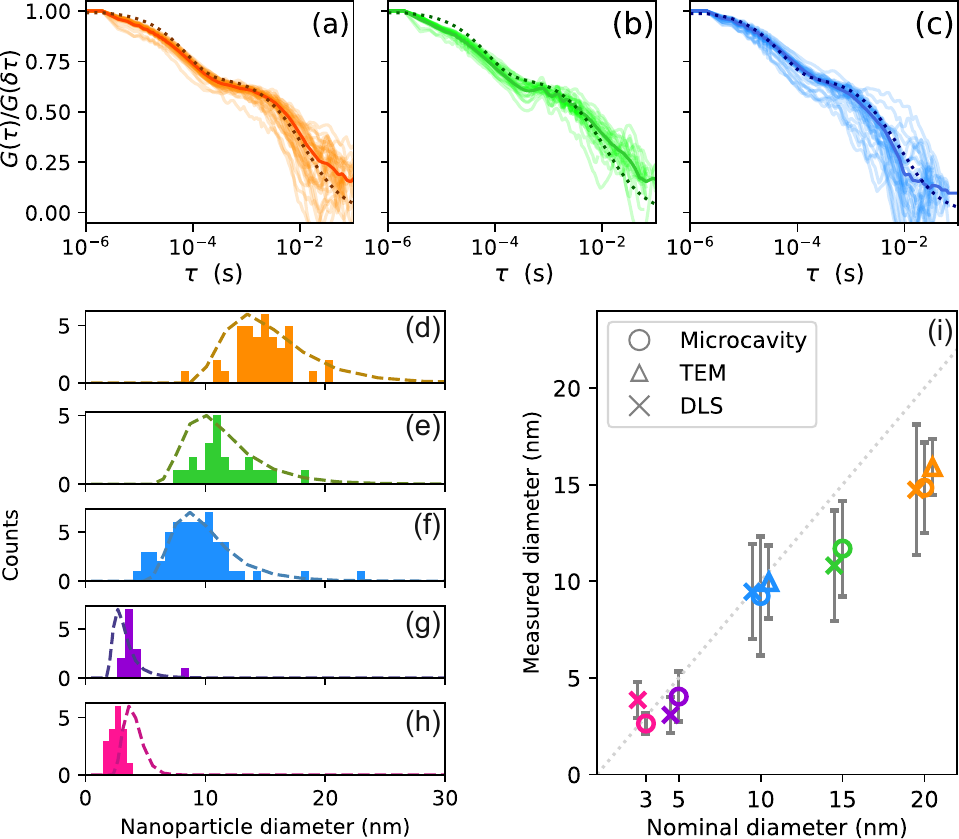}
    \caption{(a)--(c) Autocorrelations of single particle events with nanoparticles having nominal diameters 20 nm, 15 nm and 10 nm respectively. The averaged curves are shown darker, and the theoretical ACFs with the mean measured sizes are shown dotted. (d)--(h) Particle sizes measured with nanoparticles having nominal diameters diameters 20 nm, 15 nm, 10 nm, 5 nm and 3 nm respectively. For comparison, the size distributions obtained from a commercial DLS device are shown as dashed lines. (i) Comparison of nanoparticle sizing methods: the microcavity, DLS and TEM (10 nm and 20 nm only). Note that for clarity, the data points for different measurement methods are slightly horizontally offset.}
    \label{fig:nanoparticle_sizes}
\end{figure*}

\subsection{Autocorrelation of single particle events}
\label{sec:single_particles}
We now turn to the aspect of treating a transient signal with possible contributions from the lock, which requires additional considerations and signal processing.

In the first step, the offset of the lock set point and the influence of small external disturbances or of the locking mechanism in long particle events are removed by baseline-correcting the particle signal (see Supporting Information).

Second, a single nanoparticle event occupies only a part of a typical measurement, and we wish to select only this segment for calculation of the ACF. For larger particles, this region of interest can be identified by monitoring the amplitude of fluctuations above the background noise level; however, for smaller particles ($R<$ 10 nm), this is not reliable due to the small signal-to-noise ratio they produce. Therefore, we calculate $G(\delta\tau)$ as introduced above and use this value as a measure of particle presence in the trace segment, similar to the approach of Asgari \textit{et al}~\cite{asgari}. Segments with a low correlation amplitude are excluded from further analysis. It can be seen from Fig.~\ref{fig:autocorrelation}~(c) and (d) that segments exhibiting higher correlation than the background can thus be effectively identified and extracted for further processing. This method for automatic nanoparticle recognition reduces bias towards larger particles or agglomerates compared to amplitude-based recognition, and allows the extraction of particle signals with fluctuation amplitudes comparable to the noise background.

Third, having identified a nanoparticle event in the measured signal, the transient nature of a single-particle event must be taken into account. The analysis of a fluctuation signal as described earlier is valid for a stationary system, in which the fluctuations in particle concentration are small compared to the average concentration in the observation volume. In the regime of a single-particle measurement, however, the average concentration of particles in the observation volume lies in ${0<\langle C(t)\rangle<1}$, depending on the individual particle trajectory. As a result, one cannot arbitrarily subtract the mean signal level to obtain the fluctuation signal. Instead, we note from Fig.~\ref{fig:gaussian_train}~(c) that the ACF derived for a standing wave shows an initial decay corresponding to diffusion through a single antinode of the optical field and a second decay corresponding to motion between antinodes, separated by a point of minimum slope. We empirically find that the height of the point of minimum slope depends only on the cavity geometry and not on the sample. We can thus use this point as a reference level to which the amplitude of the first diffusion decay is normalized. This is done by subtracting an offset $y_0$ from the measured signal time trace, which effectively corresponds to $\langle F(t)\rangle$.

In Fig.~\ref{fig:nanoparticle_sizes}~(a)--(c), we show measured ACFs of single particle events from particles with nominal diameters of 20~nm, 15~nm, and 10~nm, respectively.
We fit the thus obtained experimental ACFs with the analytical model introduced above over the extent of the first diffusion decay (Fig.~\ref{fig:autocorrelation}~(b)). The second decay is less reliable due to the smaller statistics it represents. We use $D_T$ as the only fit parameter, which is directly related to the hydrodynamic radius $R$ via the Stokes-Einstein relation. Hence, we can extract the size of the single particle from each event. Exemplary fits are shown for the average ACFs, and a fit to a single particle signal is shown in Fig.~\ref{fig:autocorrelation}~(b). We emphasize the importance of our careful treatment by noting that using the ACF for a single Gaussian as in FCS would lead to systematic errors and increased uncertainty (see Supporting Information, Fig. S2 and Table S1).

In this manner, we measured the diameters of several single nanoparticles with specified sizes down to 3~nm and plot the obtained sizes in histograms, see Fig.~\ref{fig:nanoparticle_sizes}.

To verify the accuracy of our method, we compare our results to measurements with a commercial DLS system and TEM images, which are in excellent agreement as shown in Fig.~\ref{fig:nanoparticle_sizes}~(d),~(e). The deviations of the measured sizes from the nominal values point towards systematic uncertainties in the nominal values given by the manufacturer. It can also be observed that for the smallest two particle sizes, the cavity measurement reproduces the right size trend and is very close to the nominal diameter, while DLS shows an opposing trend, pointing towards the limitation of the latter method. Also, we note an intrinsic ambiguity in DLS results, since the scattered intensity due to a particle with radius $R$ in an ensemble measurement is proportional to $R^6$. As a result, the directly measured intensity-normalized distributions are highly vulnerable to subpopulations of larger particles or agglomerates. Instead, we have considered number-normalized distributions which, while offering a more meaningful comparison in this work, are calculated using assumptions about some nanoparticle properties. They are thus dependent on \textit{a priori} knowledge of the nanoparticles and are susceptible to errors in the provided data. Finally, in contrast to DLS measurements, only microscopic sample volumes of a few hundred microliters with ultra-low concentration are required for the microcavity experiments.

With the TEM measurements of $N=34$ and 20 individual GNPs of size 10 nm and 20 nm respectively, we compare our results to another single-particle technique, again with very good agreement, as shown in Fig. \ref{fig:nanoparticle_sizes} (i). Notably, the width of the size distribution obtained from cavity measurements matches the one from TEM measurements and is more narrow than the one from DLS. Also, the mean values of the particle size agree very closely. This indicates a high precision and accuracy of our method. Other precision single-particle sizing measurements in the literature, such as nanoparticle tracking analysis (NTA) and interferometric NTA (iNTA) experiments, result in similar distributions. \cite{kashnakova2022} We note, however, that our measurements enable sizing of more than three times smaller particles.

\section{Conclusions}

In this work, we have demonstrated a technique for measuring and analyzing single, unlabeled, freely diffusing nanoparticles down to a diameter of 3~nm, which corresponds to about 20 gold atoms across and an effective molecular weight of 160~kDa. We note that by operating far away from the plasmon resonance wavelength of gold nanospheres ($\approx$~500~nm), we can neglect plasmon enhancement, and the GNPs behave essentially like dielectric particles. Our smallest GNPs have $\Re(\alpha)$ equivalent to a silica sphere with a diameter of about 8~nm. This approaches the single-molecule sensitivity recently reported with a photothermally enhanced optical microcavity \cite{needham}, while additionally offering the capability for precise and accurate size determination. Using the modified autocorrelation model described in this work, it is possible to quantitatively extract information about single particles from the autocorrelation of their event traces without the need for \textit{a priori} calibration measurements.

Furthermore, the high measurement bandwidth possible with this method provides a temporal resolution orders of magnitude better than imaging-based nanoscopy techniques \cite{kazaian}, and can enable not only the resolution of the motion of smaller and thus faster particles, but also of fast dynamics such as rotation or internal motion such as conformational dynamics in the future.

\section{Methods}

\textbf{Experimental setup.} 
The cavity fibers were machined in-house with single-mode fibers according to the procedure developed by Hunger \textit{et al.} \cite{hunger} and the profiles had radii of curvature of 43~µm and 41~µm. The mirror coating was done by Laseroptik GmbH. The cavity is probed with a grating-stabilized diode laser at 780~nm. Transmitted light detected by an avalanche photodetector is monitored on an oscilloscope and used as the error signal for cavity locking.

For cavity locking, we implement a slow stabilization scheme designed such that rapid resonance shifts due to interaction with a diffusing nanoparticle lie outside the locking bandwidth, pass out of the cavity unattenuated and are detected in the transmission signal. Slower acoustic and mechanical jitter as well as thermal drifts are, on the other hand, corrected by the locking system. A detailed verification of this is shown in the Supporting Information, see Figs. S3 and S4.

Cavity stabilization was carried out by a piezo-electric transducer (PZT), called PZT 2, which is attached to one cavity fiber and driven by a proportional-integral-derivative (PID) feedback signal $\xi(t)$ (see Fig.~\ref{fig:setup}). Since the travel range of the locking PZT is limited by the PID controller output to about 100 nm, another PZT (PZT 1) attached to the other cavity fiber receives a d.c. offset from an arbitrary function generator (AFG) with a larger output range, which is used to bring a cavity resonance into the range of the lock PZT.
%% Theoretical travel = 3 µm with 320 V --> 200 nm with 20 V, actually less (50%) --> 100 nm

\textbf{Reference measurements.}
Reference nanoparticle measurements by DLS were performed with a Malvern Instruments Zetasizer Nano ZS equipped with a 633 nm laser, in disposable cuvettes. 
TEM images of 10 nm and 20 nm nanoparticles were acquired with a FEI Osiris electron microscope at 200 kV, with nanoparticles deposited on a carbon-film-coated copper grid. Size analysis was performed with ImageJ.

\textbf{Nanoparticle samples.}
The gold nanospheres were purchased from Nanopartz Inc. and are dispersed in a buffer of citrate in distilled water. They have a specified polydispersity of 25\% (diameter 3~nm), 20\% (5~nm) and 10\% (10~nm, 15~nm, 20~nm). The stock suspension was diluted in an isotonic citrate buffer in distilled and filtered water as recommended by the manufacturer to a concentration of about 50~pM, at which less than one particle resides in the mode volume on average at any time.

\textbf{Data processing.}
Data measured by the oscilloscope were smoothed with a Savinsky-Golay filter and resampled to 500 kHz (5~nm - 20~nm nanospheres) or 1 MHz (3~nm nanospheres). Frequency filtering was performed to remove sharp peaks arising from noise sources. The data were then processed as described above by a custom-written Python script. The autocorrelation was computed using the multiple-tau algorithm \cite{Mueller2012}. For details about the data processing steps, see Supporting Information.

\subsection{Acknowledgments}
We acknowledge L. Flatten for helpful discussions and J. Treptow for performing TEM measurements. D.H. and S.P. acknowledge funding from the Max Planck School of Photonics (MPSP) and the Karlsruhe School of Optics and Photonics (KSOP). D.H., S.P., C.R. and C.F. acknowledge funding from the Deutsche Forschungsgemeinschaft (DFG,
German Research Foundation) through the Collaborative Research Centre “4f for Future” (CRC 1573 project number 471424360, project C4).

\subsection{Competing interests}
The authors declare no competing interests.

\subsection{Author contributions}
S.P., L.K. and D.H. conceived the idea, S.P. and L.K. prepared the experimental set-up, S.P. performed microcavity experiments. S.P. and D.H. developed and performed the analysis. C.R. and C.F. performed DLS and TEM measurements. S.P. and D.H. wrote the manuscript, L.K., C.R. and C.F. provided feedback.

\subsection{Supporting information}
The following supporting information is available online: Simulated particle events and comparison of models, Simulated cavity disturbance, Signal processing, Table of results and GNP specifications.

\bibliography{references.bib}

\providecommand{\latin}[1]{#1}
\makeatletter
\providecommand{\doi}
  {\begingroup\let\do\@makeother\dospecials
  \catcode`\{=1 \catcode`\}=2 \doi@aux}
\providecommand{\doi@aux}[1]{\endgroup\texttt{#1}}
\makeatother
\providecommand*\mcitethebibliography{\thebibliography}
\csname @ifundefined\endcsname{endmcitethebibliography}  {\let\endmcitethebibliography\endthebibliography}{}
\begin{mcitethebibliography}{25}
\providecommand*\natexlab[1]{#1}
\providecommand*\mciteSetBstSublistMode[1]{}
\providecommand*\mciteSetBstMaxWidthForm[2]{}
\providecommand*\mciteBstWouldAddEndPuncttrue
  {\def\EndOfBibitem{\unskip.}}
\providecommand*\mciteBstWouldAddEndPunctfalse
  {\let\EndOfBibitem\relax}
\providecommand*\mciteSetBstMidEndSepPunct[3]{}
\providecommand*\mciteSetBstSublistLabelBeginEnd[3]{}
\providecommand*\EndOfBibitem{}
\mciteSetBstSublistMode{f}
\mciteSetBstMaxWidthForm{subitem}{(\alph{mcitesubitemcount})}
\mciteSetBstSublistLabelBeginEnd
  {\mcitemaxwidthsubitemform\space}
  {\relax}
  {\relax}

\bibitem[Taylor and Sandoghdar(2019)Taylor, and Sandoghdar]{Taylor2019}
Taylor,~R.~W.; Sandoghdar,~V. Interferometric {Scattering} {Microscopy}: {Seeing} {Single} {Nanoparticles} and {Molecules} via {Rayleigh} {Scattering}. \emph{Nano Letters} \textbf{2019}, \emph{19}, 4827--4835, Publisher: American Chemical Society\relax
\mciteBstWouldAddEndPuncttrue
\mciteSetBstMidEndSepPunct{\mcitedefaultmidpunct}
{\mcitedefaultendpunct}{\mcitedefaultseppunct}\relax
\EndOfBibitem
\bibitem[Ginsberg \latin{et~al.}(2025)Ginsberg, Hsieh, Kukura, Piliarik, and Sandoghdar]{Ginsberg2025}
Ginsberg,~N.~S.; Hsieh,~C.-L.; Kukura,~P.; Piliarik,~M.; Sandoghdar,~V. Interferometric scattering microscopy. \emph{Nature Reviews Methods Primers} \textbf{2025}, \emph{5}, 1--21, Publisher: Nature Publishing Group\relax
\mciteBstWouldAddEndPuncttrue
\mciteSetBstMidEndSepPunct{\mcitedefaultmidpunct}
{\mcitedefaultendpunct}{\mcitedefaultseppunct}\relax
\EndOfBibitem
\bibitem[Piliarik and Sandoghdar(2014)Piliarik, and Sandoghdar]{Piliarik2014}
Piliarik,~M.; Sandoghdar,~V. Direct optical sensing of single unlabelled proteins and super-resolution imaging of their binding sites. \emph{Nature Communications} \textbf{2014}, \emph{5}, 4495, Publisher: Nature Publishing Group\relax
\mciteBstWouldAddEndPuncttrue
\mciteSetBstMidEndSepPunct{\mcitedefaultmidpunct}
{\mcitedefaultendpunct}{\mcitedefaultseppunct}\relax
\EndOfBibitem
\bibitem[Dahmardeh \latin{et~al.}(2023)Dahmardeh, Dasterji, Mazal, Köstler, and Sandoghdar]{dahmardeh}
Dahmardeh,~M.; Dasterji,~H.~M.; Mazal,~H.; Köstler,~H.; Sandoghdar,~V. Self-supervised machine learning pushes the sensitivity limit in label-free detection of single proteins below 10 kDa. \emph{Nature Methods} \textbf{2023}, \emph{20}, 442--447\relax
\mciteBstWouldAddEndPuncttrue
\mciteSetBstMidEndSepPunct{\mcitedefaultmidpunct}
{\mcitedefaultendpunct}{\mcitedefaultseppunct}\relax
\EndOfBibitem
\bibitem[Kasaian \latin{et~al.}(2024)Kasaian, Mazaheri, and Sandoghdar]{kazaian}
Kasaian,~K.; Mazaheri,~M.; Sandoghdar,~V. Long-Range Three-Dimensional Tracking of Nanoparticles Using Interferometric Scattering Microscopy. \emph{ACS Nano} \textbf{2024}, \emph{18}, 30463--30472\relax
\mciteBstWouldAddEndPuncttrue
\mciteSetBstMidEndSepPunct{\mcitedefaultmidpunct}
{\mcitedefaultendpunct}{\mcitedefaultseppunct}\relax
\EndOfBibitem
\bibitem[Kashkanove \latin{et~al.}(2022)Kashkanove, Blessing, Gemeinhardt, Soulat, and Sandoghdar]{kashnakova2022}
Kashkanove,~A.~D.; Blessing,~M.; Gemeinhardt,~A.; Soulat,~D.; Sandoghdar,~V. Precision size and refractive index analysis of weakly scattering nanoparticles in polydispersions. \emph{Nature Methods} \textbf{2022}, \emph{19}, 586–593\relax
\mciteBstWouldAddEndPuncttrue
\mciteSetBstMidEndSepPunct{\mcitedefaultmidpunct}
{\mcitedefaultendpunct}{\mcitedefaultseppunct}\relax
\EndOfBibitem
\bibitem[Baaske \latin{et~al.}(2020)Baaske, Neu, and Orrit]{Baaske2020}
Baaske,~M.~D.; Neu,~P.~S.; Orrit,~M. Label-Free Plasmonic Detection of Untethered Nanometer-Sized Brownian Particles. \emph{ACS Nano} \textbf{2020}, \emph{14}, 14212--14218\relax
\mciteBstWouldAddEndPuncttrue
\mciteSetBstMidEndSepPunct{\mcitedefaultmidpunct}
{\mcitedefaultendpunct}{\mcitedefaultseppunct}\relax
\EndOfBibitem
\bibitem[Asgari \latin{et~al.}(2023)Asgari, Baaske, and Orrit]{Asgari2023}
Asgari,~N.; Baaske,~M.~D.; Orrit,~M. Burst-by-{Burst} {Measurement} of {Rotational} {Diffusion} at {Nanosecond} {Resolution} {Reveals} {Hot}-{Brownian} {Motion} and {Single}-{Chain} {Binding}. \emph{ACS Nano} \textbf{2023}, \emph{17}, 12684--12692, Publisher: American Chemical Society\relax
\mciteBstWouldAddEndPuncttrue
\mciteSetBstMidEndSepPunct{\mcitedefaultmidpunct}
{\mcitedefaultendpunct}{\mcitedefaultseppunct}\relax
\EndOfBibitem
\bibitem[Vollmer and Yang(2012)Vollmer, and Yang]{Vollmer2012}
Vollmer,~F.; Yang,~L. Label-free detection with high-{Q} microcavities: a review of biosensing mechanisms for integrated devices. \emph{Nanophotonics} \textbf{2012}, \emph{1}, 267\relax
\mciteBstWouldAddEndPuncttrue
\mciteSetBstMidEndSepPunct{\mcitedefaultmidpunct}
{\mcitedefaultendpunct}{\mcitedefaultseppunct}\relax
\EndOfBibitem
\bibitem[Foreman \latin{et~al.}(2015)Foreman, Swaim, and Vollmer]{Foreman2015}
Foreman,~M.~R.; Swaim,~J.~D.; Vollmer,~F. Whispering gallery mode sensors. \emph{Adv. Opt. Photon.} \textbf{2015}, \emph{7}, 168--240\relax
\mciteBstWouldAddEndPuncttrue
\mciteSetBstMidEndSepPunct{\mcitedefaultmidpunct}
{\mcitedefaultendpunct}{\mcitedefaultseppunct}\relax
\EndOfBibitem
\bibitem[Zhu \latin{et~al.}(2010)Zhu, Ozdemir, Xiao, Li, He, and Chen]{zhu2010}
Zhu,~J.; Ozdemir,~S.~K.; Xiao,~Y.-F.; Li,~L.; He,~L.; Chen,~L.,~Da-Renand~Yang On-chip single nanoparticle detection and sizing by mode splitting in an ultrahigh-Q microresonator. \emph{Nature Photon} \textbf{2010}, \emph{4}, 46--49\relax
\mciteBstWouldAddEndPuncttrue
\mciteSetBstMidEndSepPunct{\mcitedefaultmidpunct}
{\mcitedefaultendpunct}{\mcitedefaultseppunct}\relax
\EndOfBibitem
\bibitem[Baaske \latin{et~al.}(2014)Baaske, Foreman, and Vollmer]{Baaske2014}
Baaske,~M.~D.; Foreman,~M.~R.; Vollmer,~F. Single-molecule nucleic acid interactions monitored on a label-free microcavity biosensor platform. \emph{Nature Nanotechnology} \textbf{2014}, \emph{9}, 933--939, Publisher: Nature Publishing Group\relax
\mciteBstWouldAddEndPuncttrue
\mciteSetBstMidEndSepPunct{\mcitedefaultmidpunct}
{\mcitedefaultendpunct}{\mcitedefaultseppunct}\relax
\EndOfBibitem
\bibitem[Trichet \latin{et~al.}(2016)Trichet, Dolan, James, Hughes, Vallance, and Smith]{Trichet2016}
Trichet,~A. A.~P.; Dolan,~P.~R.; James,~D.; Hughes,~G.~M.; Vallance,~C.; Smith,~J.~M. Nanoparticle Trapping and Characterization Using Open Microcavities. \emph{Nano Letters} \textbf{2016}, \emph{16}, 6172--6177\relax
\mciteBstWouldAddEndPuncttrue
\mciteSetBstMidEndSepPunct{\mcitedefaultmidpunct}
{\mcitedefaultendpunct}{\mcitedefaultseppunct}\relax
\EndOfBibitem
\bibitem[Malmir \latin{et~al.}(2022)Malmir, Okell, Trichet, and Smith]{Smith2022}
Malmir,~K.; Okell,~W.; Trichet,~A. A.~P.; Smith,~J.~M. Characterization of nanoparticle size distributions using a microfluidic device with integrated optical microcavities. \emph{Lab Chip} \textbf{2022}, \emph{22}, 3499--3507\relax
\mciteBstWouldAddEndPuncttrue
\mciteSetBstMidEndSepPunct{\mcitedefaultmidpunct}
{\mcitedefaultendpunct}{\mcitedefaultseppunct}\relax
\EndOfBibitem
\bibitem[Kohler \latin{et~al.}(2021)Kohler, Mader, Kern, Wegener, and Hunger]{Kohler2021}
Kohler,~L.; Mader,~M.; Kern,~C.; Wegener,~M.; Hunger,~D. Tracking Brownian motion in three dimensions and characterization of individual nanoparticles using a fiber-based high-finesse microcavity. \emph{Nature Communications} \textbf{2021}, \emph{12}, 6385\relax
\mciteBstWouldAddEndPuncttrue
\mciteSetBstMidEndSepPunct{\mcitedefaultmidpunct}
{\mcitedefaultendpunct}{\mcitedefaultseppunct}\relax
\EndOfBibitem
\bibitem[Malmir \latin{et~al.}(2022)Malmir, Okell, Trichet, and Smith]{Malmir2022}
Malmir,~K.; Okell,~W.; Trichet,~A. A.~P.; Smith,~J.~M. Characterization of nanoparticle size distributions using a microfluidic device with integrated optical microcavities. \emph{Lab on a Chip} \textbf{2022}, \emph{22}, 3499--3507, Publisher: The Royal Society of Chemistry\relax
\mciteBstWouldAddEndPuncttrue
\mciteSetBstMidEndSepPunct{\mcitedefaultmidpunct}
{\mcitedefaultendpunct}{\mcitedefaultseppunct}\relax
\EndOfBibitem
\bibitem[Needham \latin{et~al.}(2024)Needham, Saavedra, Rasch, Sole-Barber, Schweitzer, Fairhall, Vollbrecht, Wan, Podorova, Bergsten, Mehlenbacher, Zhang, Tenbrake, Saimi, Kneely, Kirkwood, Pfeifer, Chapman, and Goldsmith]{needham}
Needham,~L.-M. \latin{et~al.}  Label-free detection and profiling of individual solution-phase molecules. \emph{Nature} \textbf{2024}, \emph{629}, 1062--1068\relax
\mciteBstWouldAddEndPuncttrue
\mciteSetBstMidEndSepPunct{\mcitedefaultmidpunct}
{\mcitedefaultendpunct}{\mcitedefaultseppunct}\relax
\EndOfBibitem
\bibitem[Wohland \latin{et~al.}(2020)Wohland, Machan, and Maiti]{wohland}
Wohland,~T.; Machan,~R.; Maiti,~S. \emph{An Introduction to Fluorescence Correlation Spectroscopy}; IOP Publishing, 2020\relax
\mciteBstWouldAddEndPuncttrue
\mciteSetBstMidEndSepPunct{\mcitedefaultmidpunct}
{\mcitedefaultendpunct}{\mcitedefaultseppunct}\relax
\EndOfBibitem
\bibitem[Hunger \latin{et~al.}(2010)Hunger, Steinmetz, Colombe, Deutsch, Hänsch, and Reichel]{hunger}
Hunger,~D.; Steinmetz,~T.; Colombe,~Y.; Deutsch,~C.; Hänsch,~T.~W.; Reichel,~J. A fiber Fabry–Perot cavity with high finesse. \emph{New Journal of Physics} \textbf{2010}, \emph{12}, 65038\relax
\mciteBstWouldAddEndPuncttrue
\mciteSetBstMidEndSepPunct{\mcitedefaultmidpunct}
{\mcitedefaultendpunct}{\mcitedefaultseppunct}\relax
\EndOfBibitem
\bibitem[Ries and Schwille(2012)Ries, and Schwille]{Ries2012}
Ries,~J.; Schwille,~P. Fluorescence correlation spectroscopy. \emph{BioEssays} \textbf{2012}, \emph{34}, 361--368\relax
\mciteBstWouldAddEndPuncttrue
\mciteSetBstMidEndSepPunct{\mcitedefaultmidpunct}
{\mcitedefaultendpunct}{\mcitedefaultseppunct}\relax
\EndOfBibitem
\bibitem[Elson and Magde(1974)Elson, and Magde]{elson}
Elson,~E.~L.; Magde,~D. Fluorescence correlation spectroscopy. I. Conceptual basis and theory. \emph{Biopolymers} \textbf{1974}, \emph{13}, 1--27\relax
\mciteBstWouldAddEndPuncttrue
\mciteSetBstMidEndSepPunct{\mcitedefaultmidpunct}
{\mcitedefaultendpunct}{\mcitedefaultseppunct}\relax
\EndOfBibitem
\bibitem[Dominguez-Medina \latin{et~al.}(2016)Dominguez-Medina, Chen, Blankenburg, Swanglap, Landes, and Link]{Dominguez2016}
Dominguez-Medina,~S.; Chen,~S.; Blankenburg,~J.; Swanglap,~P.; Landes,~C.~F.; Link,~S. Measuring the {Hydrodynamic} {Size} of {Nanoparticles} {Using} {Fluctuation} {Correlation} {Spectroscopy}. \emph{Annual Review of Physical Chemistry} \textbf{2016}, \emph{67}, 489--514, Publisher: Annual Reviews\relax
\mciteBstWouldAddEndPuncttrue
\mciteSetBstMidEndSepPunct{\mcitedefaultmidpunct}
{\mcitedefaultendpunct}{\mcitedefaultseppunct}\relax
\EndOfBibitem
\bibitem[Asgari \latin{et~al.}(2024)Asgari, Baaske, Ton, and Orrit]{asgari}
Asgari,~N.; Baaske,~M.~D.; Ton,~J.; Orrit,~M. Exploring Rotational Diffusion with Plasmonic Coupling. \emph{ACS Photonics} \textbf{2024}, \emph{11}, 634--641\relax
\mciteBstWouldAddEndPuncttrue
\mciteSetBstMidEndSepPunct{\mcitedefaultmidpunct}
{\mcitedefaultendpunct}{\mcitedefaultseppunct}\relax
\EndOfBibitem
\bibitem[Müller(2012)]{Mueller2012}
Müller,~P. \emph{Python multiple-tau algorithm 0.3.3}; {Python Package Index (PyPI)}, 2012\relax
\mciteBstWouldAddEndPuncttrue
\mciteSetBstMidEndSepPunct{\mcitedefaultmidpunct}
{\mcitedefaultendpunct}{\mcitedefaultseppunct}\relax
\EndOfBibitem
\end{mcitethebibliography}

% \begin{figure*}[p]
%         \includegraphics[height=4cm]{ForTableOfContentsOnly.pdf}
%         \caption{For table of contents only.}
% \end{figure*}

\end{document}

\renewcommand{\figurename}{Figure}
\renewcommand{\thefigure}{S\arabic{figure}}

\renewcommand{\tablename}{Table}
\renewcommand{\thetable}{S\arabic{table}}

\section{Simulated particle events and comparison of models}
\label{sec:simulations}

\begin{figure}[h!]
    \centering
        \includegraphics[width=1\textwidth]{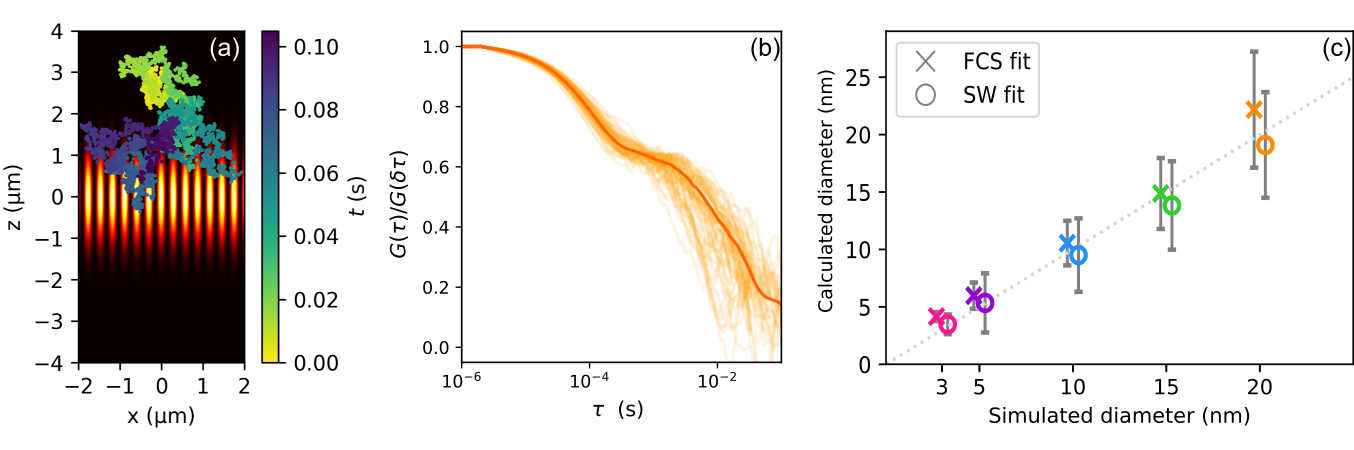}
    \caption{\hl{(a) Illustration of a simulated diffusing particle's path through the ccavity field. (b) Autocorrelations of 100 simulated nanoparticle events with a 20 nm nanospheres. The averaged autocorrelation is shown darker. (c) Comparison between the calculated diameters obtained from the FCS and the SW models.}}
    \label{fig:simulations}
\end{figure}

\hl{To verify the accuracy of our measurement technique and model, we performed simulations of diffusing nanoparticles in the cavity field approximated by the Gaussian series as described in the main text. Performing identical analyses on the simulated data as described in the main text, we obtain results closely matching the simulated particle diameters. We further compared the effect of using $G_\mathrm{FCS}(\tau)$ and $G_\mathrm{SW}(\tau)$ for fitting (Figure }\ref{fig:simulations}\hl{), and observe an overestimation by the former, which becomes more prominent for smaller particles, most likely because only a small region of the ACF at small $\tau$ can be used for this fit.}

\begin{table}[h!]
\centering
\begin{tabular}{|c|c|cc|}
\hline
{Sample} & Simulated & \multicolumn{2}{c|}{Relative deviation}                                          \\
\cline{3-4} 
\rule{0pt}{2pt}
no.                           &  diameter (nm)                                & \multicolumn{1}{c|}{of FCS fit} & \multicolumn{1}{c|}{of SW fit}\\ 
\hline
1                           & 3                                        & \multicolumn{1}{c|}{38\%}    & \multicolumn{1}{c|}{16\%}\\ 
2                           & 5                                        & \multicolumn{1}{c|}{19\%}    & \multicolumn{1}{c|}{7\%}\\ 
3                           & 10                                       & \multicolumn{1}{c|}{5\%}    & \multicolumn{1}{c|}{-5\%} \\ 
4                           & 15                                       & \multicolumn{1}{c|}{-1\%}       & \multicolumn{1}{c|}{-8\%}\\
5                           & 20                                       & \multicolumn{1}{c|}{11\%}       & \multicolumn{1}{c|}{-4\%} \\ \hline
\end{tabular}
\caption{\hl{The relative deviation of average results from nanoparticle simulations analysed using the FCS model and the SW model.}}
\end{table}
\label{tab:fit_comparison}

\begin{figure}[h!]
    \centering
        \includegraphics[width=0.65\textwidth]{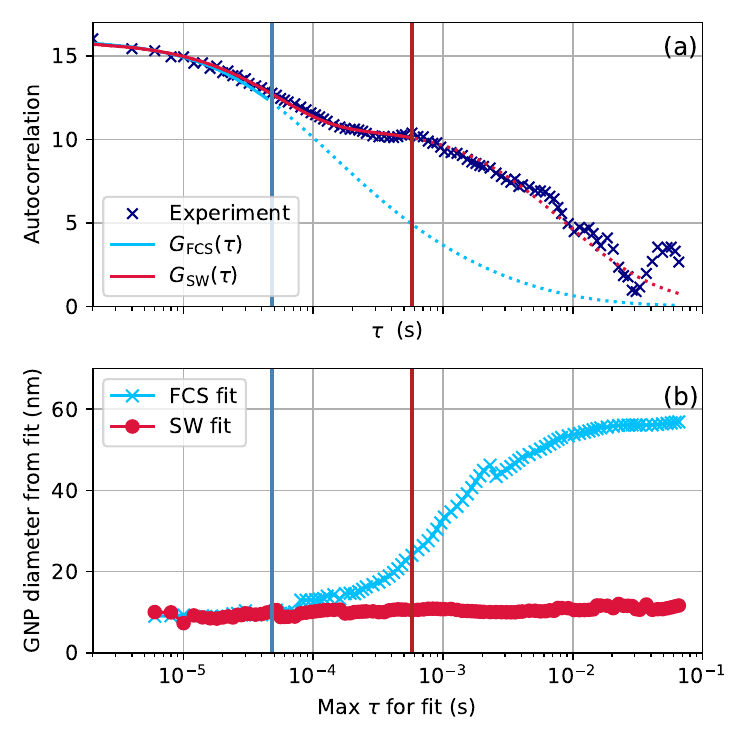}
    \caption{\hl{(a) Fitting the autocorrelation of a single nanosphere event with the ACFs from the FCS model and the SW model. The vertical lines show the respective fit limits used for evaluation. (b) Variation of diameter determined from the autocorrelation using different fit limits. The SW model is robust and gives a constant output over a large fitting range, whereas the FCS model soon deviates from the experimental data after $\tau\approx6\times10^{-5}$ s.}}
    \label{fig:fit_limits}
\end{figure}

\hl{We also compare regions of validity of the two models for fitting the autocorrelation function. In Figure }\ref{fig:fit_limits}\hl{ (b), we see that the nanoparticle diameter output by the fit to $G_\mathrm{SW}(\tau)$ remains nearly constant with respect to the extent of the autocorrelation fit, whereas that from $G_\mathrm{FCS}(\tau)$ soon diverges once $\tau\approx6\times10^{-5}$ s.}

\section{Simulated cavity disturbance}
\label{simulated_disturbance}
\begin{figure}[h]
    \centering
        \includegraphics[width=0.8\textwidth]{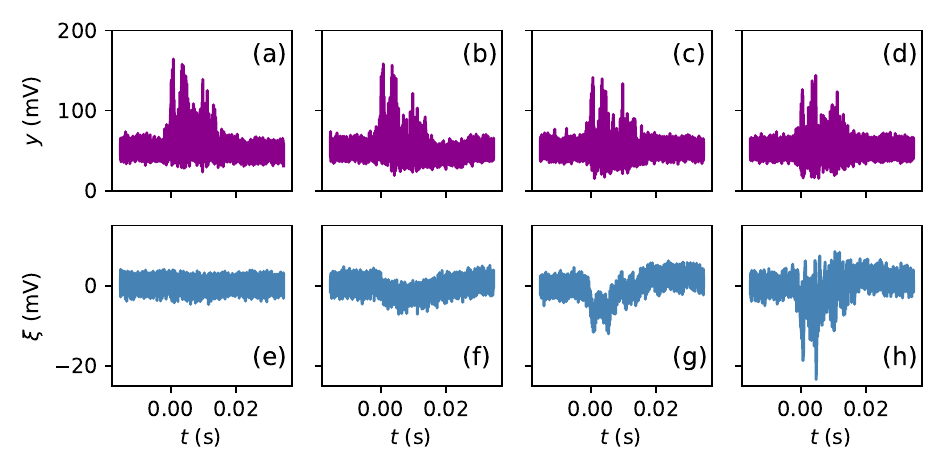}
    \caption{(a)-(d): Cavity transmission signal $y(t)$ with virtual nanoparticle modulation with unity-gain frequencies (UGFs) of 80 Hz, 800 Hz, 8 kHz, 80 kHz respectively. (e)-(h): The lock feedback signal $\xi(t)$ for the corresponding UGFs.}
    \label{fig:virtual_signals}
\end{figure}

To verify that the nanoparticle signal passes through without attenuation or filtering, we investigated the response of the lock to a simulated particle-like disturbance in the cavity. The signal due to the diffusion of a nanoparticle in an optical standing wave was simulated as described above and played back to PZT 1 by the AFG, leading to a disturbance in physical cavity length which simulates the dispersive disturbance due to a diffusing nanoparticle. The cavity transmission signal and lock signal were monitored for different lock parameters, and it was determined that for the configuration used, the nanoparticle disturbance can be fully measured in the cavity transmission signal. The influence of the unity-gain frequency (UGF), i.e. the lock bandwidth, on the cavity transmission and the lock signal is shown in Figure~\ref{fig:virtual_signals}. 

\begin{figure}[h]
    \centering
        \includegraphics[width=0.8\textwidth]{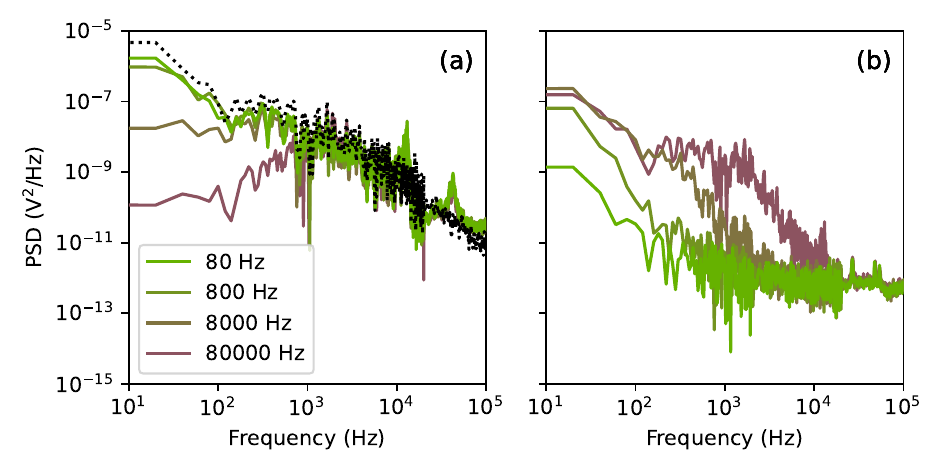}
    \caption{(a) Power spectral density (PSD) of the cavity transmission with virtual nanoparticle modulation, with varying UGFs. The PSD of the nanoparticle signal itself is shown by the black dotted line. (b) PSD of the lock feedback signal for the same UGFs shown in (a).}
    \label{fig:virtual_spectra}
\end{figure}

In Fig. \ref{fig:virtual_spectra}, the power spectral density (PSD) of the transmission signal and the lock signal are compared for different lock parameters, showing that the full bandwidth of the nanoparticle signal (shown as a dotted black spectrum) passes through to the detected cavity transmission for UGF = 80 Hz, which was used in our experiments. The close agreement between the spectra of the disturbance and the transmission also confirm that the cavity response to a nanoparticle is not affected by photothermal effects at the optical power levels used.

At higher locking bandwidths, the lock corrects the nanoparticle disturbance as well, which is seen as low-frequency attenuation of the PSD of the cavity transmission and high-pass filtering of the signal itself. When the lock bandwidth is reduced further, the stability of the cavity decreases, making it unsuitable for nanoparticle detection. The root-mean-square stability of the cavity length is around 300~fm (corresponding to a frequency jitter of 33 MHz) in the underwater cavity used in our experiments.

\section{Signal processing}
\label{signal_processing}

In this section, we outline the signal processing steps undertaken prior to autocorrelation. The cavity transmission signal is processed as follows:

\begin{enumerate}
     \item The lower envelope and mean of the signal is subtracted to yield a signal $y_0(t)$ centred at zero,
     \item The autocorrelation amplitude, $G(\delta\tau)$ is computed for 10-ms-long segments of the signal and segments featuring a nanoparticle event are hence identified,
     \item The autocorrelation of the nanoparticle signal and its level at the point of minimum slope between the two characteristic decays are computed. 
     \item An offset $Y_0$ is numerically computed such that the minimum slope level of the autocorrelation of $y_0(t)-Y_0$ matches that of the theoretical ACF calculated for the cavity geometry used (see main text).
\end{enumerate}
 
With the raw data thus preprocessed, the particle size can be determined by fitting the analytical ACF to the experimental autocorrelation.

\begin{figure}[h]
    \centering
        \includegraphics[width=1.0\textwidth]{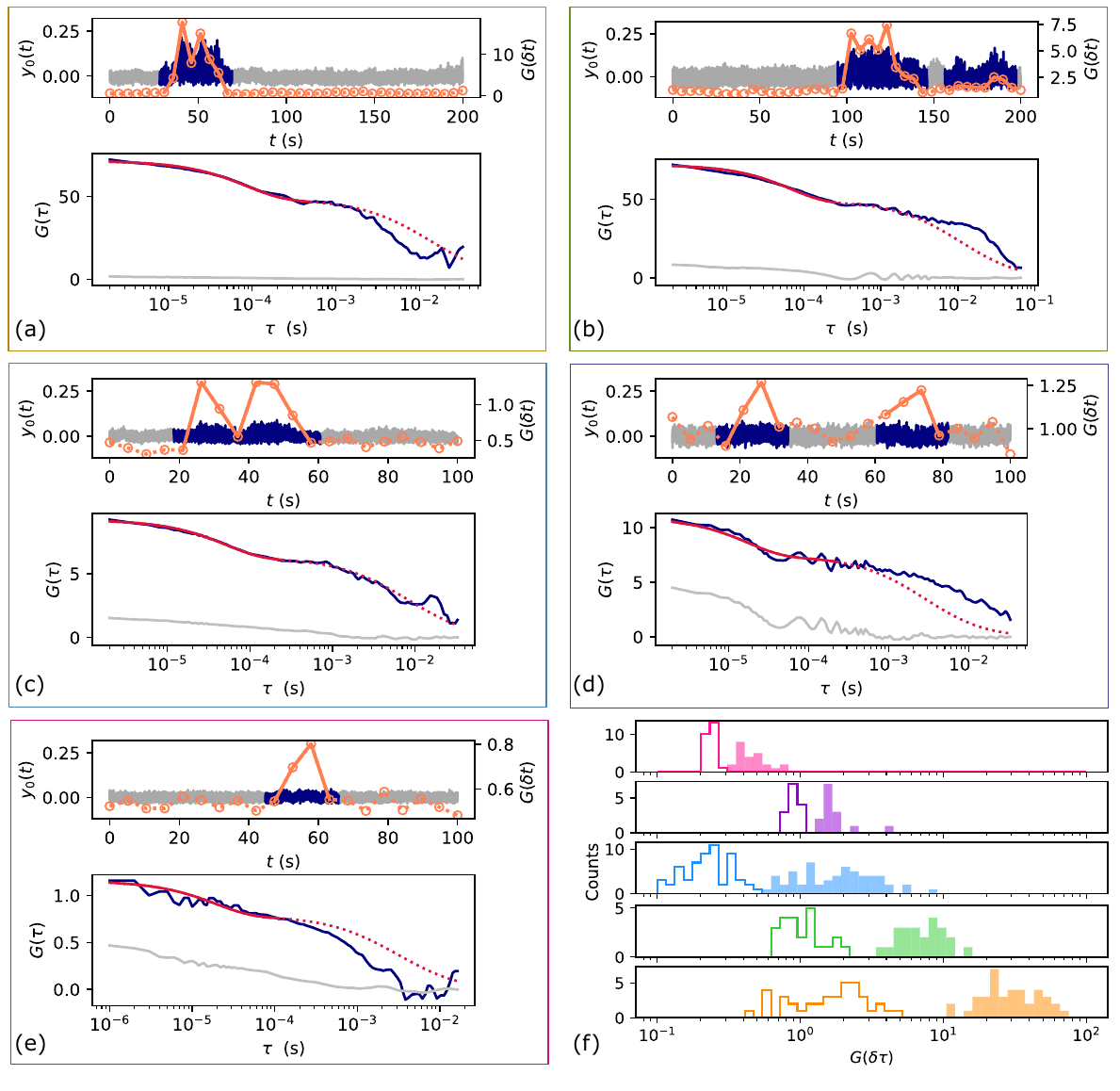}
    \caption{(a) - (e) Top panel: Measured signals separated into nanoparticle (dark blue) and background (grey) segments based on the autocorrelation amplitude $G(t; \delta\tau)$ of 5 ms segments (orange). Bottom panel: Autocorrelation of the nanoparticle event (dark blue) and background section (grey), with the fit of $G_\mathrm{SW}(\tau)$ in red. The figures correspond to particles with nominal sizes of 20 nm, 15 nm, 10 nm, 5 nm, 3 nm. (f) Distribution of the maximum autocorrelation amplitudes $G(\delta\tau)$ of a nanoparticle event (solid histograms) and the background (unfilled histograms) for 3~nm, 5~nm, 10~nm, 15~nm and 20~nm GNPs. The experiments with 3~nm and 10~nm showed a lower background noise than the others.}
    \label{fig:signal_processing}
\end{figure}

Examples of measured nanoparticle events and $G(\delta\tau)$ are shown in Figure \ref{fig:signal_processing} (a) - (e), along with their autocorrelation. 

We note that at the smallest particles measured in this work, the correlation approaches the background correlation level, introducing a lower limit to the size of particles which can currently be detected with our setup.
 
\section{Table of results and GNP specifications}
\label{results_table}

In the following table, we summarize the results of microcavity measurements as well as comparison measurements performed by dynamic light scattering (DLS) and transmission electron microscopy (TEM). The DLS results are shown number-normalised.

\begin{table}[h!]
\centering
\begin{tabular}{|c|c|ccc|}
\hline
{Sample} & Specified & \multicolumn{3}{c|}{Measured diameter (nm)}                                          \\
\cline{3-5} 
\rule{0pt}{2pt}
no.                           &  diameter (nm)                                & \multicolumn{1}{c|}{with microcavity} & \multicolumn{1}{c|}{by DLS}         & by TEM\\ 
\hline
1                           & 3                                        & \multicolumn{1}{c|}{2.6 $\pm$ 0.5}    & \multicolumn{1}{c|}{3.8 $\pm$ 0.8}  & -  \\ 
2                           & 5                                        & \multicolumn{1}{c|}{4.0 $\pm$ 1.3}    & \multicolumn{1}{c|}{3.1 $\pm$ 0.9}  & -  \\ 
3                           & 10                                       & \multicolumn{1}{c|}{9 $\pm$ 3}    & \multicolumn{1}{c|}{9.5 $\pm$ 2.5}  & 10 $\pm$ 2  \\ 
4                           & 15                                       & \multicolumn{1}{c|}{12 $\pm$ 2}       & \multicolumn{1}{c|}{11 $\pm$ 3} & -  \\
5                           & 20                                       & \multicolumn{1}{c|}{15 $\pm$ 2}       & \multicolumn{1}{c|}{15 $\pm$ 3} & 15.9 $\pm$ 1.4 \\ \hline
\end{tabular}
\caption{A comparison of nanosphere diameters measured using our microcavity and with a commercial DLS system. The DLS results are number-normalised. The microcavity results represent statistics from $N =$ 15, 18, 53, 22 and 37 single particle measurements for sample numbers 1, 2, 3, 4 and 5 respectively.}
\end{table}
\label{tab:diameters}

\hl{In the following table, we list the specified nanoparticle sizes and uncertainties given by the manufacturer.}

\begin{table}[h!]
\centering
\begin{tabular}{|c|c|c|c|c|}
\hline

\hline
Sample no.                  & Nominal (nm)                             & Spec. low (nm)         & Spec. high (nm)    & PDI  \\ \hline
1                           & 3                                        & 2                     & 4                  & 25\%  \\ 
2                           & 5                                        & 4                     & 7                  & 20\%  \\ 
3                           & 10                                       & 8                     & 12                 & 10\%  \\ 
4                           & 15                                       & 13                    & 17                 & 10\%  \\
5                           & 20                                       & 18                    & 22                 & 10\% \\ \hline
\end{tabular}
\caption{Specifications of the nanoparticle manufacturer (Nanopartz, Inc.) for the GNP samples investigated. PDI: polydispersity index.}
\end{table}
\label{tab:specifications}